\documentstyle[12pt]{article}
\begin{document}
\begin{titlepage}
\pagestyle{empty}
\baselineskip=21pt
\rightline{Alberta-THY-2/94}
\rightline{UMN-TH-1237/94}
\rightline{hep-ph/9402294}
\rightline{January 1994}
\vskip .2in
\begin{center}
{\large{\bf Neutrino Mass Effects in a Minimally Extended
Supersymmetric Standard Model}} \end{center}
\vskip .1in
\begin{center}
Rouzbeh Allahverdi
  and
Bruce A. Campbell

{\it Department of Physics, University of Alberta}

{\it Edmonton, Alberta, Canada T6G 2J1}

and

Keith A. Olive

{\it School of Physics and Astronomy, University of Minnesota}

{\it Minneapolis, MN 55455, USA}

\vskip .1in

\end{center}
\vskip .5in
\centerline{ {\bf Abstract} }
\baselineskip=18pt
We consider an extension of the supersymmetric standard model which
includes
singlet Higgs superfield representations (in three generations) to
generate
neutrino masses
 via the see-saw
mechanism. The resulting theory may then exhibit R-parity violation
in the couplings of the singlets, inducing $R$-parity violating
effective
interactions
among the standard model superfields, as well as inducing decay of
the lightest
neutralino,
which otherwise would compose a stable LSP. We compute the rates for
the
resulting
neutralino decays, depending on the particular superpotential
couplings
responsible for the
violation of R-parity. We compare to astrophysical constraints on the
decay of
massive particles.

\noindent
\end{titlepage}
\baselineskip=18pt

\def\la{~\mbox{\raisebox{-.7ex}{$\stackrel{<}{\sim}$}}~}
\def\ga{~\mbox{\raisebox{-.7ex}{$\stackrel{>}{\sim}$}}~}
\def\mtw#1{m_{\tilde #1}}
\def\tw#1{${\tilde #1}$}
\def\beq{\begin{equation}}
\def\eeq{\end{equation}}

Despite its current experimental elusiveness, the supersymmetric
standard model
remains theoretically well motivated. Supersymmetry itself
represents the
unique
possibility to combine internal and spacetime symmetries in quantum
field
theories,
evading ``no go" theorems by its incorporation of anticommutation
relations in
its
defining algebraic structure.  The gauging of supersymmetry
inevitably results
in
a generally coordinate invariant theory, and Einsteinian gravity, and
leads
along
the path of unification of gravity with the gauge interactions.
Unification of
the gauge
interactions themselves appears to be facilitated by supersymmetry;
extrapolation of
the gauge coupling constants of the standard model gauge group
factors
according
to the renormalization group equation with standard model matter
content does
not
result in them coming together at a single point. On the other hand,
when the
superpartners
of the standard model matter multiplets (including two Higgs doublets
as
required in the
supersymmetric standard model) are included in the renormalization
group
running
above the electroweak scale then the coupling constants for SU(3),
SU(2), and
U(1),
cross at an energy scale of order $10^{16}$ GeV \cite{amaldi},
 as would be required
for
unification,
and this unification scale is consistent, in these theories, with the
observed
stability
of the proton. Finally, the inclusion of  supersymmetry partners at
about the
electroweak
scale is essential for the strongest phenomenological motivation for
supersymmetry, which
is to explain the stability of the electroweak scale under radiative
corrections, and the
maintenance of the hierarchy between the electroweak scale and the
GUT or
Planck
scales.

It is well known, however, that the minimal supersymmetric standard
model
(MSSM)
contains only a
few of the possible gauge invariant couplings. With a minimal field
content,
the
superpotential can be written as the sum of Yukawa terms (with
generation
indices
suppressed)
\beq
F_Y = h_u H_1 Q u^c + h_d H_2 Q d^c + h_e H_2 L e^c
\eeq
and to avoid an axion like state a mass term mixing the two Higgs
doublets $H_1$ and $H_2$,
\beq
F_H = \mu  H_1 H_2
\eeq
$Q$ and $L$ are weak doublets and $u^c, d^c$, and $ e^c$ are their
corresponding
right-handed  counterparts.
These are the only superpotential terms necessary to recover standard
model
fermion masses and Higgs couplings. The MSSM possesses a $Z_2$
symmetry known
as
$R$-parity which can be represented by $R = (-1)^{F + L + 3B}$ in
terms of the
particle's fermion, lepton, and baryon numbers.

One obvious extension of the MSSM consists of the inclusion of
neutrino masses via a see-saw mechanism\cite{seesaw}.  This is easily
accomplished by the
addition to the superpotential of
\beq
F_\nu = M N N + h_\nu H_1 L N
\eeq
Neutrino masses of order ${h_\nu}^2 {v_1}^2/ M$ will then be
generated for the
light
(left-handed) neutrinos, where $v_1 = \langle H_1 \rangle$.
It should be noted that the interactions induced by the
superpotential $F_\nu$
do not
violate R-parity as they they only violate lepton number in units of
two. They
do
not, however, constitute the most general set of N-field interactions
allowed
by gauge
invariance. As well as $F_\nu$, one may also introduce the
superpotential terms
\beq
F_N = \lambda H_1 H_2 N + k N^3
\label{fn}
\eeq
The combination of the interactions in $F_\nu$ with {\em either} of
the
interactions
in $F_N$ will result in violation of $R$-parity.

The inclusion of neutrino mass by the see-saw mechanism
 has many other benefits in addition to the generation
of neutrino masses which can in principle aid in the solution to the
solar
neutrino
problem and/or atmospheric neutrino deficit and/or cosmological hot
dark matter
(though not all simultaneously without the inclusion of a fourth
sterile
neutrino).
Right-handed neutrino decay has been utilized\cite{fy1} to generate a
lepton
asymmetry which
in conjunction with non-perturbative electroweak interactions becomes
a baryon
asymmetry.
In refs. \cite{cdo2,mur}, this mechanism was extended to
supersymmetric models
as well.
Another possibility\cite{cdo1} for the generation of a baryon
asymmetry made
use of flat directions
in the scalar potential as in the  the Affleck-Dine
mechanism\cite{ad}.
In the latter, the superpotential $F = F_Y + F_N + F_\nu$ was
required in order
to
induce a lepton
number violating operator. For simplicity only one set (3
generations) of
chiral superfields
were added. Thus $R$-parity was explicitly violated. In that model,
$R$-parity
could have been preserved if the $N$-fields in $F_N$ were distinct
and have a
different $R$-parity assignment than that of the $N$'s in $F_\nu$.

There are numerous other ways in which one can imagine extending the
MSSM.
In what is often called the minimal-nonminimal superymmetric standard
model
(MNMSSM)
a single additional gauge singlet chiral superfield, $N$ is
added\cite{N}. This
extension is realized by simply adding to the superpotential the
contribution
from
$F_N$ (\ref{fn}).
 The primary motivation for the inclusion of the Higgs singlet is the
possibility that
it offers for the dynamical generation of the Higgs mixing mass
$\mu$. If  the
$N$ field
is a field which acquires a vev determined by mass parameters
of the order of the electroweak scale, then with a $NH_{1}H_{2}$
coupling of
standard strength
(say comparable to a gauge coupling) Higgs mixing of the requisite
magnitude is
induced.
On the other hand, if the mass parameters in the $N$ sector are much
larger,
say of an
intermediate scale, or perhaps of the GUT scale, as might
naturally be
expected to be
in see-saw models, then if the $N$ has a nonzero vev one would
naturally expect
it to
also be of this scale. In such a case one still might imagine
inducing a weak
scale mixing
between the Higgs doublets, at the price of fine tuning the
$NH_{1}H_{2}$
coupling to
be small to give the hierarchical ratio between the electroweak scale
and the
$N$ mass scale.
Though this small (O($M_W$)  mixing mass is technically feasible its
smallness
is
part and parcel with the hierarchy problem. The cubic term
is required in order to avoid an $N$-axion like field, in the absence
of an
explicit
$\mu $ superpotential term mixing the two Higgs supermultiplets. In a
detailed
exaimation of this
model\cite{eghrz}, it was found that many of the standard Higgs mass
relations
are
altered.  If the MSSM Higgs mass  relations are found to be
experimentally not
viable,
this model becomes the simplest alternative.

{}From another point of view, the MNMSSM is of interest as it can
easily
produce a relatively
light dark matter candidate \cite{fot}. In the MSSM, steadily
improving
accelerator limits,
are pushing up the mass of the lightest supersymmetric particle
(LSP), which
due to the
unbroken $R$-parity in the MSSM is stable.  In the minimal model the
LSP is
generally
expected to be a linear combination of the four neutral $R = -1$
fermions\cite{ehnos}, the two
gauginos,  \tw B and \tw W, and the Higgsinos \tw {H_1} and \tw
{H_2}.  With
regards to a
dark matter candidate, the best choice in the MSSM appears to be the
bino whose
mass is typically
between 40 GeV and  $\sim 300$ GeV for cosmologically interesting
parameters\cite{susydm}.
In the non-minimal model it is quite feasible\cite{fot,sar1} to have
a a light
LSP (10 - 50 GeV),
which is a state which has a strong admixture the fernionic component
of $N$.
Though, cosmologically, a very massive LSP is just as good as a light
one
(light still referring
to O(GeV)), from the point of view of experimental detection, the
lighter one
is better\cite{fot2}.

In this letter we derive the consequences of the $R$-parity violation
of the
full superpotential.
$R$-parity violation in the quark sector is usually avoided in order
to insure
a relatively stable
proton.  In the Higgs-lepton sector, there are many constraints
on $R$-parity violation as well.  In the case we consider here,
$R$-parity
is violated only in the heavy $N$-field sector.
Nevertheless, this $R$-parity violation shows up in the low energy
sector,
most notably in the destabilization of the LSP.  We derive
constraints on the
neutrino mass parameters as a consequence of the constraints on
late-decaying
LSP's.

As well as the destabilization of the LSP to which we will turn
below, there
are
other possible low-energy signatures of  R-parity violation in the
high energy
N-field sector. If supersymmetry were exact, then even the combined
presence
of the $F_{\nu}$ and $F_N$ superpotential terms would not induce
(super)renormalizable
lepton number violating superpotential terms involving only the light
superfields of
the theory, due to the nonrenormalization theorems for the
superpotential.
After
supersymmetry breaking the nonrenormalization theorems no longer hold
exactly,
and
lepton number (and hence R-parity)violating effective interactions
will be
induced
in an amount governed by the scale of supersymmetry breaking.
This will result in low energy R-parity violating
interactions
involving standard model superfields of the form of both induced
effective
superpotential
terms such as
\beq
F_{RX} = m_{X}H_{1}L + \lambda_{X} LLe^{c}
\eeq
as well as soft supersymmetry breaking lepton number violating terms.
By
appropriate
change of basis we may diagonalize the Higgs-lepton mass mixing and
parametrize
our lepton number violating effects by $\lambda_{X}$. These terms are
induced
from
one loop diagrams in amounts
\beq
\lambda_{X}\sim{m_{\delta}^{2} \over M_{N}^{2}} ~~~~~~~~~~~~
or~~~~~~~~~~~\lambda_{X}\sim{{\mu }m_{\delta} \over M_{N}^{2}}
\eeq
where $m_\delta$ is the scale of supersymmetry breaking.
Lepton number violating renormalizable interactions  of  this type
are
constrained
by laboratory limits on lepton flavour violation, and neutrinoless
double beta
decay \cite{hs}.
As we have analyzed previously, even stronger limits are imposed on
interactions of
this type by the persistence of a baryon asymmetry in the early
universe,
assuming
that it is not produced at or after the electroweak phase transition
\cite{cdeo12,sonia}.
The danger here is that the lepton number violation implied by the
new
interaction
could attain thermal equilibrium at the same time as baryon and
lepton number
violating (but B-L conserving) nonperturbative electroweak
interaction effects
to
simultaneously equilibrate both the baryon and lepton number of the
universe to
zero.
If these limits pertain, they would imply that $
\lambda_{X}<7\times10^{-7}$\cite{cdeo12}. These
limits may be evaded, and indeed a baryon asymmetry may be generated
from a
lepton asymmetry, provided one of the generations of lepton flavours
has its
lepton
number violating interaction in equilibrium, while another does not
\cite{cdeo3}.

As we have mentioned above, the combination of the $NH_{1}L$
superpotential
term
with either the $NH_{1}H_{2}$ or $NNN$ superpotential interactions
breaks
R-parity and hence will destabilize the lightest neutralino mass
eigenstate.
The nature of the resulting decay will depend on which of these
latter terms
is responsible. Let us begin our discussion with consideration of
decays
induced by the $NH_{1}H_{2}$ term. There will be tree-level two body
decays
to lepton-Higgs final states induced by the diagrams shown in figure
1(a),
and 1(b). We see that for decay from an $\tilde{H_{2}}$ component of
a
neutralino,
the vev is the large  $H_{1}$ vev, favoring that amplitude over the
amplitude
for the decay from diagram 1(a) with decay from the $\tilde{H_{1}}$
component
of the neutralino, by a factor of tan$\beta = v_1/v_2$,
 the  ratio of the vevs.
In addition there will be favoured (by tan$\beta$) decay amplitudes
for the
decays from the  $\tilde{H_{2}}$ and $\tilde{H_{1}}$
components of the neutralino coming
from figure
1(b).  To get the
approximate contribution to the decay of the amplitude of figure
1(a), we note
that
the insertion of the Higgs vev induces a mass mixing, of the
neutralino with
the N
field, of magnitude $m_{NHH}/M$, where $m_{NHH}$ is the mixing mass
$m_{NHH}=\lambda v  \sin \beta$, where $v^2 = v_1^2 + v_2^2$
 and $M$ is the
Majorana mass
term for the N-field.
The N-field component of the resulting mass eigenstate then induces a
decay to
an $H_{1}L$ final state (suppressing lepton generation indices) via
the
$h_\nu N H_{1} L$ superpotential coupling. Similarly, in diagram
1(b),
insertion of
the Higgs vev induces a mass mixing of the outgoing lepton with the
N-field,
 of magnitude $m_{\nu}/M$, where $m_{\nu}$ is the Dirac neutrino mass
$m_{\nu}=h_{\nu} v  \sin \beta$, and $M$ is the
Majorana mass
term for the N-field.
These mixings then appear in decay amplitudes induced by the coupling
at the
other vertex, into a two body final state, with decay width (ignoring
mixing
factors)
of the form:
\begin{equation}
\Gamma_{o} \simeq \frac{m_{\tilde \chi_{o}}}{16\pi } (1-
\frac{m_{o}^{2}}{m_{\tilde
\chi_{o}}^{2}})
\end{equation}
where $m_{o}$ is the mass of the final state Higgs scalar.
If (in the absence of mixing with the N-field)
we would write the LSP as an admixture
\begin{equation}
\tilde \chi_{o} = \alpha \tilde B^{o} + \beta \tilde W^{o} + \gamma
\tilde
H_{1}^{o}
  + \delta \tilde H_{2}^{o}
\end{equation}
Then the decay of the LSP via its $H_{2}$ component will then occur
at a rate
\begin{equation}
\Gamma \simeq \delta^{2} \frac{4 \lambda^2 h_\nu^2 v^2
\sin^2 \beta }{M_{N}^{2}}\Gamma_{o}
\label{g1}
\end{equation}
while there would be a contribution to the neutralino decay width
from decay of
its $H_{1}$ component with a contribution
\begin{equation}
\Gamma \simeq \gamma^{2} \frac{\lambda^2 h_\nu^2 v^2
\sin^2 \beta}{M_{N}^{2}}\Gamma_{o}
\label{g2}
\end{equation}

There will also be decays into neutrino-gamma modes induced at one
loop, as
shown in
figure 2. They will give a decay rate with the same mixing factors as
the tree
level
modes, multiplied by loop induced dipole decay width.

Now to produce a two body decay to a neutrino plus physical (on mass
shell)
photon, the only part of the electromagnetic vertex which contributes
is the
induced transition dipole piece, which we may parametrize as
\cite{leeschrock}
\begin{equation}
M_{\mu}=-i\bar{u}(p_{f}){i \sigma_{\mu\nu} q^{\nu} \over
(m_{\tilde \chi_{o}}+m_{\nu})}
(F_{2}^{V} + F_{2}^{A}\gamma_{5}) u(p_{i})
\end{equation}
which results in a dipole decay rate (dropping the neutrino mass)
\begin{equation}
\Gamma_{D}={m_{\tilde \chi_{o}} \over 8\pi}\left[ [F_{2}^{V} ]^{2}+
[F_{2}^{A}
]^{2}\right]
\end{equation}
{}From inspection of the diagram we find that the mixing factors
associated
with
the N mass and the Higgs vevs must combine with the kinematics of the
dipole
decay to give a net decay width of order
\begin{equation}
\Gamma \simeq  {m_{\tilde \chi_{o}} \over 8\pi}{v^4
\sin^4 \beta h_{\nu}^{2} \lambda^2 e^2 \over
M_{N}^2 m^2}
\end{equation}
where $m$ is a mass of the order the  electroweak scale. Note that in
order to
induce a dipole matrix element we have to have broken supersymmetry.
This implicitly appears in our estimate in that lines in the loop,
which is
dominated
by momenta of order the electroweak scale, are split in mass by
supersymmetry
breaking of order that scale, giving a result whose magnitude we may
read off
an individual diagram as above.  We also note that there is no
diagram
involving
the N-field in a loop in such a way as to induce a dipole with less
suppression
by
the N-field mass, as such diagrams involve photon emission from
external lines,
and the Ward-Takahashi identities of electromagnetic gauge invariance
ensure
that such terms do not contribute to the induced
dipole \cite{leeschrock}.

Similarly, decays of the LSP may be induced by the $NNN$
superpotential term,
as represented by the diagrams of figure 3. We note that figure 3(a)
is an
induced D-term and contains a loop which is also a D-term. This
ensures a
non-zero decay rate for the neutralino even when supersymmetry is
unbroken,
unlike the case for F-terms. Because D-terms do not obey
non-renormalization
theorems, they can be radiatively induced even in the limit of
unbroken
supersymmetry; hence in general they appear without suppression
factors
associated with the scale of supersymmetry breaking.We also note that
the
induced D-term in figure 3(a) (and its associated component diagrams)
is a
dimension six term \cite{weinberg}. The component diagrams relevant
to
neutralino decay are shown in figure 3(b) to 3(e). The processes of
figure 3
dominate
over decays induced by tree-level diagrams for large $M_{N}$, as the
latter are
suppressed by eight powers of  $M_{N}$ in rate, whereas the loop
induced decays
are only suppressed by four powers. Computing the
diagrams of
figures  3(b) and 3(c)
one finds that they result in a decay rate that is approximately
\beq
\Gamma \sim \gamma^2 {k^{2} h_{\nu}^{6}  \over 16\pi{(2\pi)}^8}
{{\mu^2}v_{1}^{2}
m_{\tilde \chi_{o}}
\over M_{N}^4}
\eeq
whereas the final two diagrams of figure 3 result in a decay rate for
the LSP
that is
approximately
\begin{equation}
\Gamma \sim \gamma^2 {k^{2} h_{\nu}^{6}  \over 16\pi{(2\pi)}^8}
{v_{1}^{2}
m_{\tilde \chi_{o}}^3 \over
M_{N}^4}
\label{3N}
\end{equation}
We expect that the Higgsino mass should be at least of the order of
the doublet
mixing
term, and in certain circumstances the doublet mixing term might be
substantially smaller;
so below we will  use the latter of these rate estimates for
numerical
estimates.


 Almost without exception, the LSP decays we are considering are
effectively
entropy producing decays, ie. they will produce high energy photons.
Photon producing decays are known to be highly constrained from both
astrophyical and cosmological observations (see eg. ref. \cite{eglns}
for a recent compilation of such limits).  These limits generally
place
constraints
in the density-lifetime plane of the decaying particle.
We will assume that the LSP, in the absense of its decay, is the
dominant form
of dark matter and therefore assume that its cosmological density is
such that
$\Omega_\chi \approx 1$, where $\Omega = \rho/\rho_c$ is the
cosmological
density
parameter. At this density, one finds that the LSP lifetime is
constrained so
that
either $\tau_\chi \la 10^4$s to avoid affecting the light element
abundances
produced
during big bang nucleosynthesis, or the LSP must be effectively
stable with a
lifetime $\tau_\chi \ga 10^{24}$s. Astophysical limits on other R-
parity violaing interactions were considered in \cite{bs}.

The decay rates in Eqs. (\ref{g1},\ref{g2},\ref{3N}) are clearly
dependent on a
number
of model parameters.  In order to get a feeling for the limits
imposed by the
cosmological constraints we make a few more assumptions.  We assume
that the
LSP is primarily a gaugino (a bino) with mass $m_\chi \approx 150$
GeV.
For $|\mu| \sim 1-10$ TeV, $\gamma \sim 2 \times 10^{-3} - 2 \times
10^{-2}$
and $\delta
\sim  4 \times 10^{-3} - 4 \times 10^{-2}$ and for large $\tan
\beta$, $\sin
\beta \approx
1$.  We can then write (for the decays based on the $H_1H_2N$
superpotential
term)
\beq
\tau_\chi \simeq 3 \times 10^{-6} h_\nu^{-2}
\lambda^{-2} (4\delta^{2}
 +  \gamma^{2})^{-1}
\left({M_N \over 10^{12} {\rm GeV}}\right)^2
\left({150 {\rm GeV} \over m_\chi}\right) {\rm s}
\eeq
Taking central values for $\gamma$ and $\delta$, and $h_\nu \sim
\lambda \sim
h$,
we have
\beq
\tau_\chi \simeq 7 \times 10^{-3} h^{-4} \left({M_N \over 10^{12}
{\rm
GeV}}\right)^2
\left({150 {\rm GeV} \over m_\chi}\right) {\rm s}
\eeq
The constraints on $\tau_\chi$ are therefore constraints on $M_N$,
\beq
M_N \la 10^{15} h^2 ~{\rm GeV}
\eeq
or
\beq
M_N \ga 10^{25} h^2 ~{\rm GeV}
\eeq
The latter limit, of course only makes sense for $h \ll 1$ and in
this case the
LSP
is effectively stable as its lifetime is much greater than the age of
the
Universe.

For LSP decay induced by the $kN^3$ superpotential term, from the
decay width
estimate given above we deduce an LSP lifetime of order
\beq
\tau_\chi \simeq  4 \times 10^{20} {h_\nu}^{-6} k^{-2} \gamma^{-2}
 \left({M_N \over 10^{12} {\rm GeV}}\right)^4
\left({150 {\rm GeV} \over m_\chi}\right)^3 {\rm s}
\eeq
which translates into the limits
\beq
M_N \la 5\times 10^{6} {h_\nu}^{3/2} k^{1/2} ~{\rm GeV}
\eeq
or
\beq
M_N \ga 5 \times 10^{11} {h_\nu}^{3/2} k^{1/2}~{\rm GeV}
\eeq

These limits show therefore that even if $R$-parity violation is
inserted
in the singlet sector, destabilization of the LSP can indeed occur
and
$R$-parity violation of this type is strongly constrained. It is
especially
interesting that cosmological arguments provide such strong
constraints,
probing possible see-saw sources of R-parity violation to far higher
mass scales than could be directly accessed by laboratory experiment;
this provides yet another example of the power of cosmological
considerations
to provide us with new information about the fundamental interactions
of
nature.

\vskip 1.0truecm
\noindent {\bf Acknowledgements}
\vskip 1.0truecm
This work was supported in part by  the Natural Sciences and
Engineering
Research Council of Canada and by  DOE grant DE-FG02-94ER40823.
The work of KAO was in addition supported by a Presidential Young
Investigator Award.

\newpage
\noindent{\bf{Figure Captions}}

\vskip.3truein

\begin{itemize}

\item[]
\begin{enumerate}
\item[]
\begin{enumerate}
\item[{\bf Figure 1:}] Neutralino decay diagrams induced by an
$NH_{1}H_{2}$
superpotential term.

\item[{\bf Figure 2:}] Radiative neutralino decay diagrams induced by
an
$NH_{1}H_{2}$ superpotential term.

\item[{\bf  Figure 3:}] Neutralino decay diagrams induced by an $NNN$
superpotential term. Figure 3(a) is the superfield diagram whose
dominant associated component field diagrams include those shown in
Figures
3(b) through 3(e).

\end{enumerate}
\end{enumerate}
\end{itemize}

\end{document}